\documentclass[A4,twocolumn]{webofc}
\usepackage[varg]{txfonts}   
\usepackage{color}

\def \pgamma {\mbox{$p \gamma$}\xspace}
\def \fpg {\mbox{$f_{p \gamma}$}\xspace}
\def \gammarayAdj {\mbox{$\gamma$}-ray\xspace}

\def \gg {\mbox {$\gamma \gamma$}\xspace}
\newcommand{\gevcmsqs}{\mbox{GeV~cm$^{-2}$~s$^{-1}$}}

\def \TXS {\mbox{TXS\,0506+056}\xspace}
\def\be{\begin{equation}}
\def\ee{\end{equation}}
\def\ba{\begin{eqnarray}}
\def \ea{\end{eqnarray}}
\def \vareps {\varepsilon}

\begin{document}
\title{High-Energy Neutrinos from Blazar Flares and Implications of TXS 0506+056}

\author{\firstname{Foteini} \lastname{Oikonomou}\inst{1}\fnsep\thanks{presenter, \email{foikonom@eso.org}}, 
\firstname{Kohta} \lastname{Murase}\inst{2,3,4,5}\fnsep\thanks{\email{murase@psu.edu}} \and
        \firstname{Maria} \lastname{Petropoulou}\inst{6}\fnsep\thanks{\email{m.petropoulou@astro.princeton.edu}}}

\institute{{European Southern Observatory, Karl-Schwarzschild-Str. 2,
    Garching bei M\"unchen D-85748, Germany} \and{Department of
    Physics, The Pennsylvania State University, University Park, PA
    16802, USA} \and{Department of Astronomy \& Astrophysics, The
    Pennsylvania State University, University Park, PA 16802, USA}
  \and{Center for Particle and Gravitational Astrophysics, The
    Pennsylvania State University, University Park, PA 16802, USA}
  \and{Center for Gravitational Physics, Yukawa Institute for
    Theoretical Physics, Kyoto University, Kyoto, Kyoto 606-8502,
    Japan} \and{Department of Astrophysical Sciences, Princeton
    University, New Jersey 08544, USA}}

\abstract{Motivated by the observation of a $> 290$~TeV muon neutrino
by IceCube, coincident with a $\sim$6 month-long $\gamma$-ray flare of the blazar TXS\,0506+056, and an archival search which revealed $13 \pm 5$ further, lower-energy neutrinos in the direction of the source in 2014-2015, we discuss the likely contribution of blazars to the diffuse high-energy neutrino intensity, the implications for neutrino emission from TXS\,0506+056 based on multi-wavelength observations of the source, and a multi-zone model that allows for sufficient neutrino emission so as to reconcile the multi-wavelength cascade constraints with the neutrino emission seen by IceCube in the direction of TXS\,0506+056.}
\maketitle
\section{Introduction}
\label{intro}
The IceCube Collaboration recently reported the observation of a
$>290$~TeV muon neutrino, IceCube-170922A, coincident with a $\sim$6
month-long $\gamma$-ray flare of the blazar TXS 0506+056
\cite{IceCube:2018dnn}, at redshift $z = 0.3365$
\cite{Paiano:2018qeq}. The neutrino detection prompted
multi-wavelength electromagnetic follow-up of the source, and the
blazar flare was observed with several instruments at
energies up to > 100 GeV \cite{Ahnen:2018mvi}. 
The correlation of the neutrino with the flare of TXS 0506+056 is inconsistent with arising by chance at the $\sim 3\sigma$ level.
An archival search revealed $13 \pm 5$
further, lower-energy, neutrinos in the direction of \TXS
during a 6-month period in 2014-2015 \cite{IceCube:2018cha}. These
events were not accompanied by a $\gamma$-ray flare. Such an
accumulation of events is inconsistent with arising from a background
fluctuation at the 3.5$\sigma$ level. 
Motivated by these observations in \cite{Murase:2018iyl}, we considered
 the implications of the possible neutrino-blazar flare association. 
Here, we summarise and update some of these results, in light of
recent IceCube and other related analysis updates.
In section \ref{subsec:sec-1} we
review existing constraints to blazars as dominant sources of the diffuse
neutrino intensity seen by IceCube. In section \ref{sec:sec2.2} we determine 
the duty cycle of $\gamma$-ray flares of several {\sl Fermi-}detected blazars 
and compute the resulting neutrino enhancement. In section \ref{subsec:sec-2} we
present the constraints imposed to single-zone models of neutrino
emission from the contemporaneous broadband spectral energy
distribution (SED) of the source, and from X-ray and \gammarayAdj
observations of putative neutrino sources in general. Section
\ref{sec:sec-3} presents a multi-zone model of neutrino emission
 based on cosmic-ray induced neutral beams, which allows for 
 enhanced neutrino production with respect to standard one-zone models, 
 without violating the multi-wavelength constraints from the SED 
 of TXS\,0506+056. Conclusions are presented in section \ref{sec:sec-4}.
\section{Blazar contribution to the diffuse neutrino flux}

\subsection{Clustering constraints}
\label{subsec:sec-1}

The absence of high-energy multiplets in the IceCube data can be used
to constrain the number density of sources contributing to the diffuse
neutrino background \cite{Murase:2016gly,Ahlers:2014ioa} (see also
earlier work by
\cite{Lipari:2008zf,Silvestri:2009xb,Murase:2012df}). The IceCube
diffuse flux, $E^2_{\nu} \Phi_{\nu}$, gives an estimate of the
neutrino production rate which is at the level of $\sim (c \cdot t_{H} \cdot n_{\rm eff} \cdot \vareps_{\nu}
L_{\vareps_{\nu}}^{\rm ave}) / \Delta \Omega $ for standard candle sources with
effective number density, $ n_{\rm eff}$ and time-averaged
luminosity $ \vareps_{\nu} L_{\vareps_{\nu}}^{\rm ave}$ \cite{Waxman:1998yy}. Here, $t_{H}$ is the Hubble time, $c$ the speed of light, and $\Delta \Omega$, the solid angle covered by the detector. Here, and throughout, primed quantities correspond to comoving frame quantities. Unprimed scripted, $\vareps_{\rm X}$, corresponds to cosmic-rest frame energy, and capital unscripted, $E_{\rm X}$, to observer frame energy. 

As shown in \cite{Murase:2016gly}, one can express the number of high-energy neutrino doublets, $N_{m \geq 2}$, as, 
\be
N_{m \geq 2} = \sqrt{\pi} q_{L} \left( \frac{\Delta \Omega}{3} \right) n_0^{\rm eff} d_{n = 1}^3, 
\label{eq:doublet}\ee
 where, $d_{n = 1}$, is the distance within which the number of
expected events from a single source is equal to 1, $n_0^{\rm eff} = n_{\rm eff} [z = 0]$
the local source number density, and $q_{L}$, is a
luminosity dependent function that depends on the redshift evolution
of the source population and approaches unity for
low-luminosities. For example, $q_L = 0.94$ and $q_L = 2.0$ for
redshift evolution of the form $n_s(z) \propto (1+z)^0$ and $n_s(z)
\propto (1+z)^3$ respectively, for luminosity corresponding to $d_{n =
  1}/(c/H_0) = 0.1$, where $H_0$ is the Hubble constant.

\noindent Using eq. \ref{eq:doublet}, and expressing $d_{n = 1}$, as
\cite{Murase:2016gly}, 
\be 
d_{n=1} = \left(\frac{\vareps_{\nu}L^{\rm
    ave}_{\vareps_{\nu}}}{4 \pi F_{\rm lim}/2.4}\right) ^{1/2}, 
\ee
where $F_{\rm lim}\sim 3 \times{10}^{-10}$~\gevcmsqs~\cite{Aartsen:2018ywr} is the IceCube
point-source sensitivity (90\% CL) for an $E^{-2}$ neutrino spectrum,
one can derive a luminosity dependent {\it upper limit} on $n_{\rm
  eff}$, in the absence of doublets in the IceCube data, i.e. imposing
$N_{m \geq 2} < 1$,
\ba 
\label{eq:n0}
n_0^{\rm eff}&\lesssim&1.9\times10^{-10}~{\rm
  Mpc^{-3}}~\left(\frac{\varepsilon_\nu L_{\varepsilon_{\nu_\mu}}^{\rm
    ave}}{{10}^{44}~{\rm
    erg\,s^{-1}}}\right)^{-3/2}\nonumber\\ &\times&{q_L}^{-1}F_{\rm
  lim,-9.5}^{3/2}\left(\frac{2\pi}{\Delta\Omega}\right), \ea where
$F_{\rm lim} = 10^{-9.5} F_{\rm lim,-9.5}~{\rm GeV}~{\rm
  cm}^{-2}~{s}^{-1}$.
Using eq. \ref{eq:n0} we can derive an 
upper limit to the contribution of a source population with number
density $n_{0, \rm eff}$ to the IceCube all-flavor neutrino flux, 
\ba E_{\nu}^2 \Phi_{\rm \nu} &\approx&  \frac{\xi_z c t_{H}}{4\pi}3\varepsilon_{\nu} L_{\varepsilon_{\nu}}^{\rm ave} n_0^{\rm eff} \nonumber \\
&\lesssim&  6.9 \times 10^{-9}~{\rm GeV}~{\rm cm}^{-2}~{\rm s}^{-1}~{\rm sr}^{-1} \nonumber \\
&\times & q_L^{-2/3} \left(\frac{\xi_z}{0.7} \right) \left(\frac{2\pi}{\Delta \Omega}\right)^{2/3} \nonumber \\ 
& \times & \left( \frac{n_0^{\rm eff}}{10^{-7}~{\rm Mpc^{-3}}}\right)^{1/3} F_{\rm lim,-9.5},\label{eq:upperLim} \ea
\noindent where the function $\xi_z$, has been introduced to parametrise the redshift evolution of the neutrino emissivity of the source population; $\xi_z = 0.7$ for the \gammarayAdj luminosity density evolution of BL Lacs, $\xi_z = 8$ for that of FSRQs \cite{Ajello:2013lka}, and $\xi_z = 3$ for the X-ray luminosity density evolution of AGNs \cite{Ueda:2014tma}. 

The above limits give conservative constraints on flaring neutrino sources, as shown in \cite{Murase:2018iyl}. In general, the multiplet limits apply to transient sources, and therefore in the present context to blazar flares. The rate density is constrained as,
\ba
\label{eq:rateLimit}
\rho^{\rm eff}_0 \gtrsim& 1.7& \times 10^{4}~{\rm Gpc}^{-3}~{\rm yr}^{-1} q_{L}^2 \left(\frac{\Delta \Omega}{2\pi}\right)^2\left(\frac{T_{\rm obs}}{\rm 8 ~yr}\right)^2 \nonumber \\ &\cdot&\left(\frac{\xi_{z}}{0.7}\right)^{-3}\phi_{\rm lim,-1}^{-3}{\rm max}[N_{\rm fl},1],\,\,\,\,\,\,\,\,\,\,\,\,\,\,\,\,\,\,\,\,\,\,\,\,\,\,
\ea
where we substituted $N_{\rm fl} \approx T_{\rm obs}/{\Delta T}_{\rm fl}$, with $N_{\rm fl}$, the number of flaring periods, and $T_{\rm obs}$, the observation duration, and $\phi_{\rm lim} = 0.1 \,\phi_{\rm lim,-1}~{\rm GeV}~{\rm cm}^{-2}$ is the muon neutrino fluence sensitivity, $\phi_{\rm lim} \sim 0.04 \,{\rm GeV}~{\rm cm}^{-2}$, which can be calculated by the publicly available effective area. The limit of equation \ref{eq:rateLimit} applies to the 2017, as well as the 2014-15 flares. Even in the latter case, where a large number of low energy neutrinos were observed, the number of $>50$~TeV neutrino multiplets, $N_{m \geq 2}\leq 1$ and the limit of equation \label{eq:rateLimit} holds.
For rare transients (no repeating bursts in the observation time), we have $\rho_0^{\rm eff} = n_0^{\rm eff}/{\Delta T}_{\rm fl}$, with ${\Delta T}_{\rm fl}$, the interval between flares. The average number of sources in flaring state is $n_0^{\rm eff}{\Delta T}_{\rm dur}/{\Delta T}_{\rm fl}$.

Based on the above arguments, we conclude that the absence of doublets and higher order multiplets in the IceCube data, disfavours BL Lacs as the dominant source of the diffuse IceCube flux.
However, the constraints could be relaxed for several reasons~\cite{Murase:2018iyl}. For example, rapidly evolving FSRQs (including MeV blazars) could make a substantial contribution only in the PeV range (see also, e.g.,~\cite{Dermer:2014vaa,Neronov:2018wuo}).

The total blazar contribution to the diffuse neutrino background has further been constrained in other analyses of event clustering and autocorrelation \cite{Aartsen:2014ivk,Aartsen:2017kru}. In addition, the absence of extremely-high energy neutrinos (>5 PeV) in diffuse searches so far \cite{Aartsen:2018vtx}, constrains optimistic models of blazar neutrino emission \citep[e.g.][]{Muecke:2002bi,Padovani:2015mba}. The reason is that typically blazar neutrino models peak at $ > $ PeV energy, as the target photon density is larger at lower energies in blazars. Finally, the contribution of blazars to the IceCube flux has been constrained by cross-correlation and stacking analyses~\cite{Aartsen:2016lir,Aartsen:2017kru}. The latest update of this analysis constrains the contribution of $\gamma$-ray bright blazar emission to $\leq 27\%$ of the IceCube flux.

\subsection{Contribution from blazar flares: Duty cycle and neutrino enhancement factor}\label{sec:sec2.2}

In order to assess the fraction of neutrinos that could be produced during 
flaring states we studied the \gammarayAdj light-curves of a sample
 of blazars from the {\it Fermi} Large Area Telescope, FAVA catalogue \cite{2017ApJ...846...34A}. 

In order to investigate the fraction of time spent in a flaring state
 we introduce $f_{\rm fl}$, the flare {\it duty factor}, defined as, 
\begin{equation}
\label{eq:fl}
f_{\rm fl}=\frac{1}{N_{\rm tot}}\int_{L^{\rm th}}{\rm d}L\, \frac{{\rm d}N}{{\rm d}L},
\end{equation}
where $N_{\rm tot}$ is the total number of time bins, $L_{\rm th}$ is the threshold
 luminosity beyond which we consider the source to be flaring,
  and ${\rm d}N/{\rm d} L$ is the distribution of luminosity states. 
\begin{figure*}
\centering
\includegraphics[width=7cm,clip]{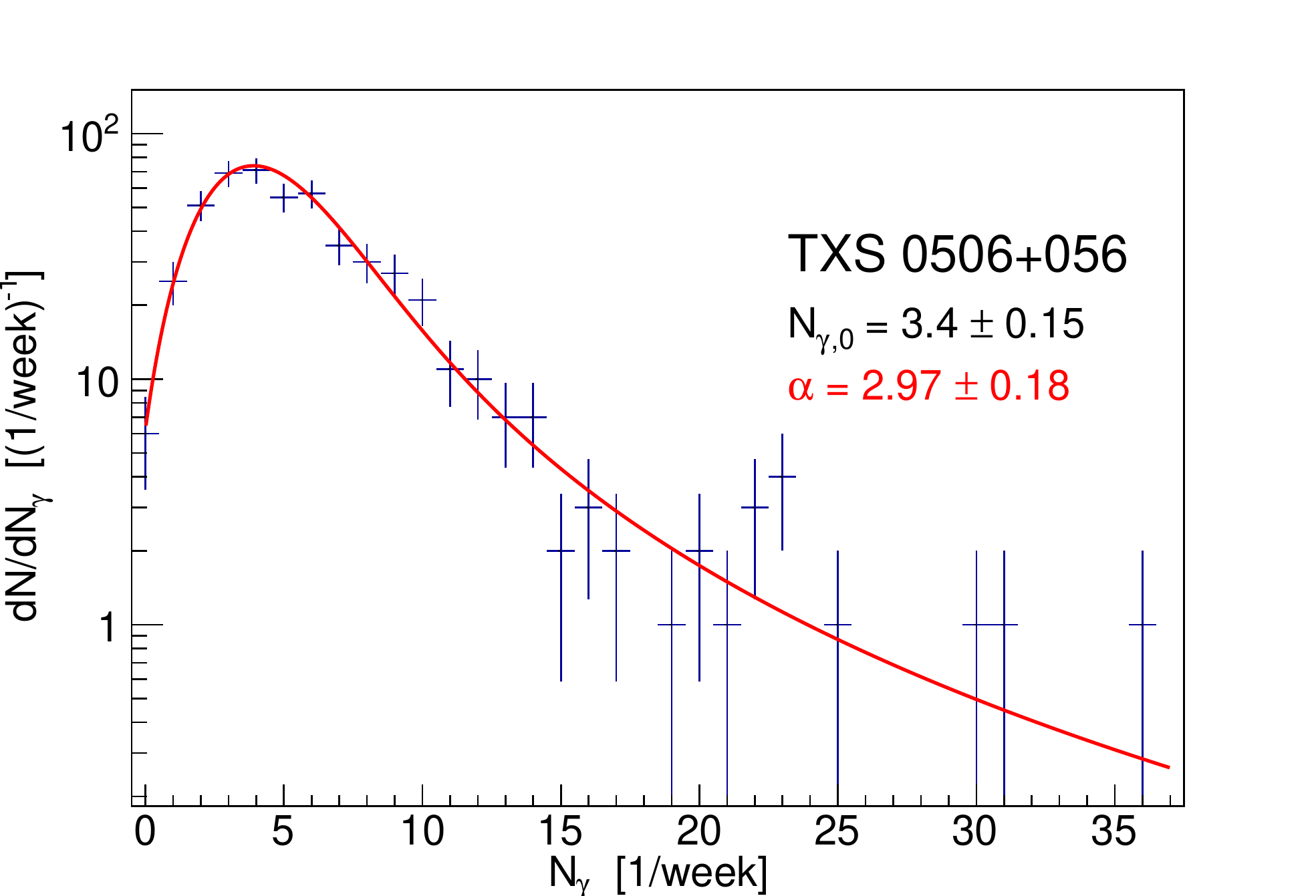}
\includegraphics[width=7cm,clip]{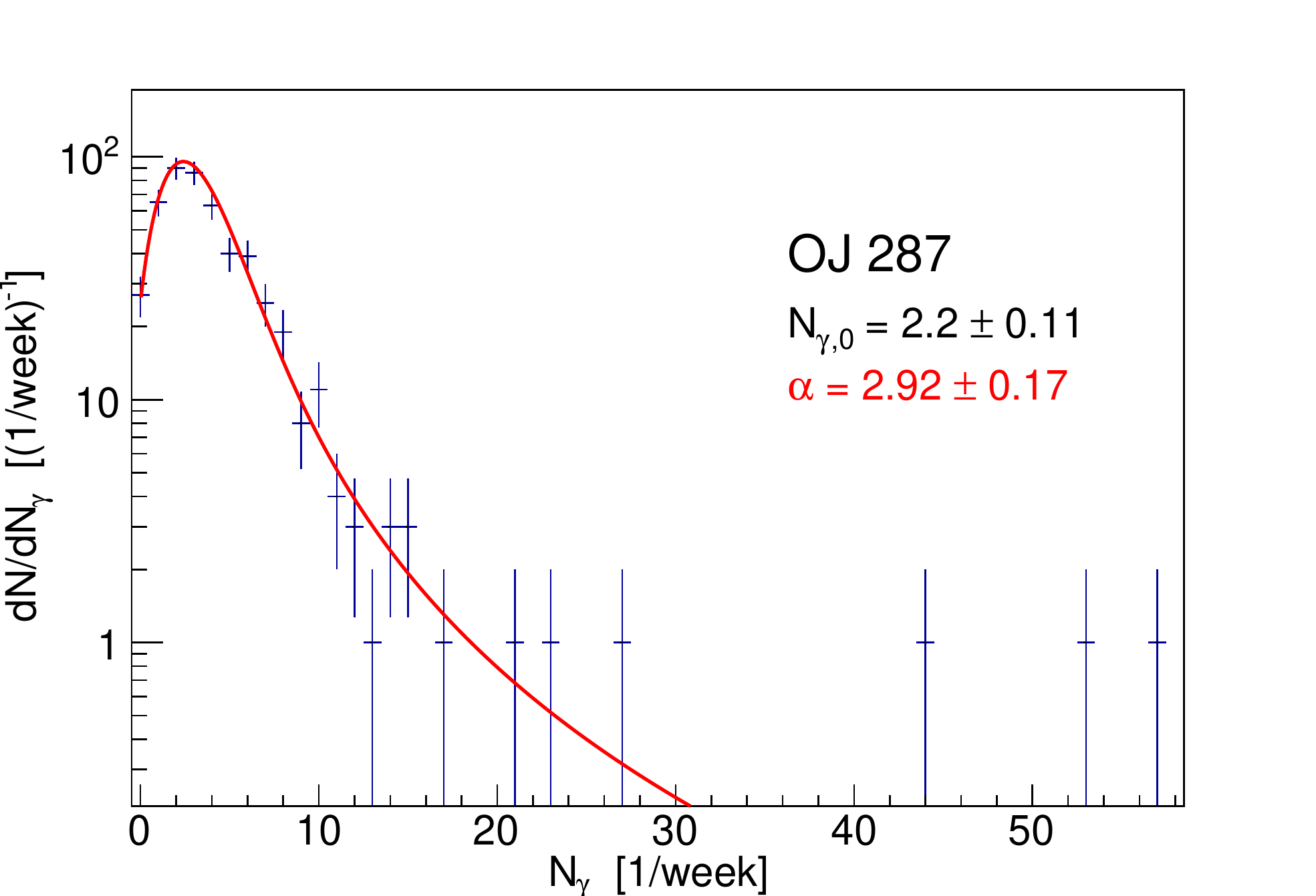}
\includegraphics[width=7cm,clip]{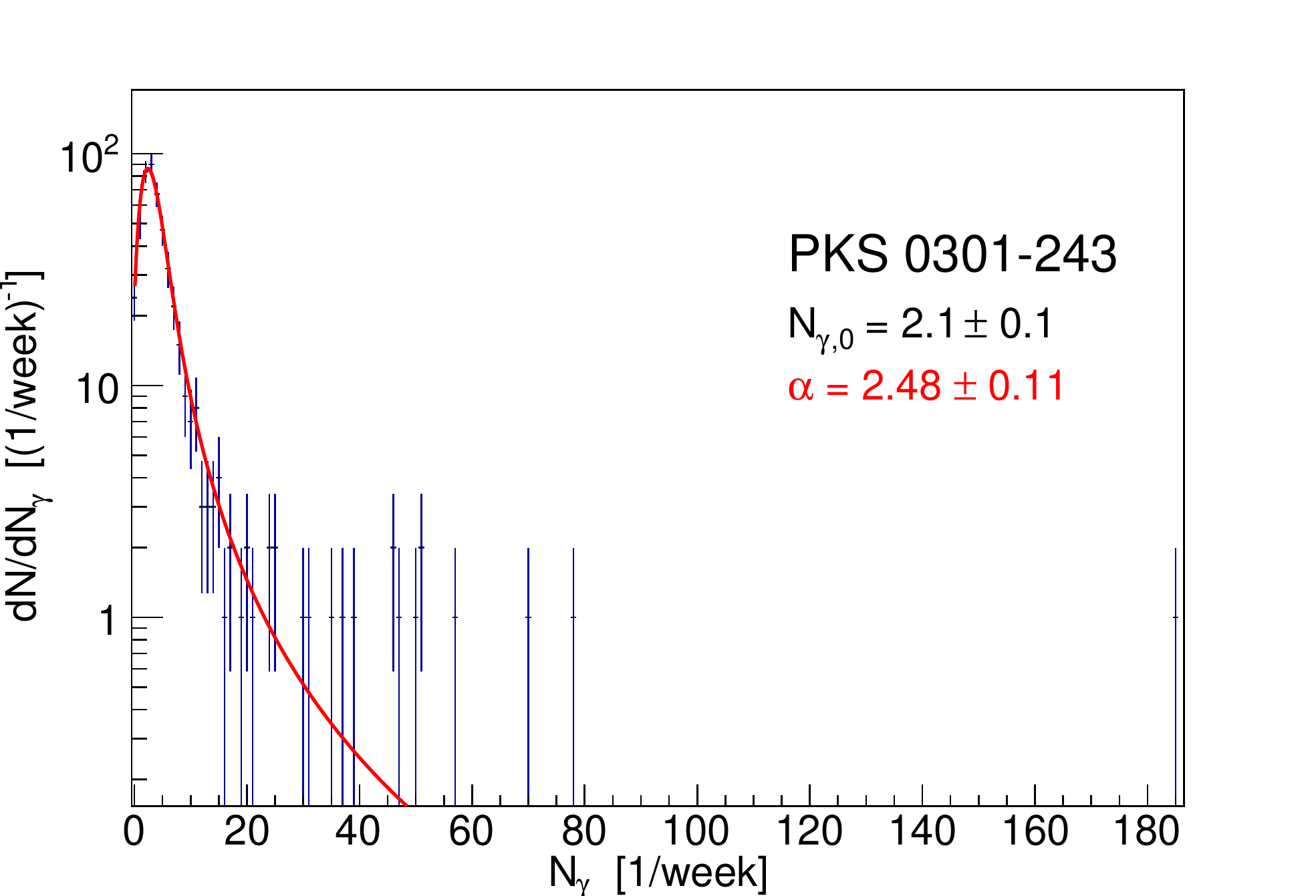}
\includegraphics[width=7cm,clip]{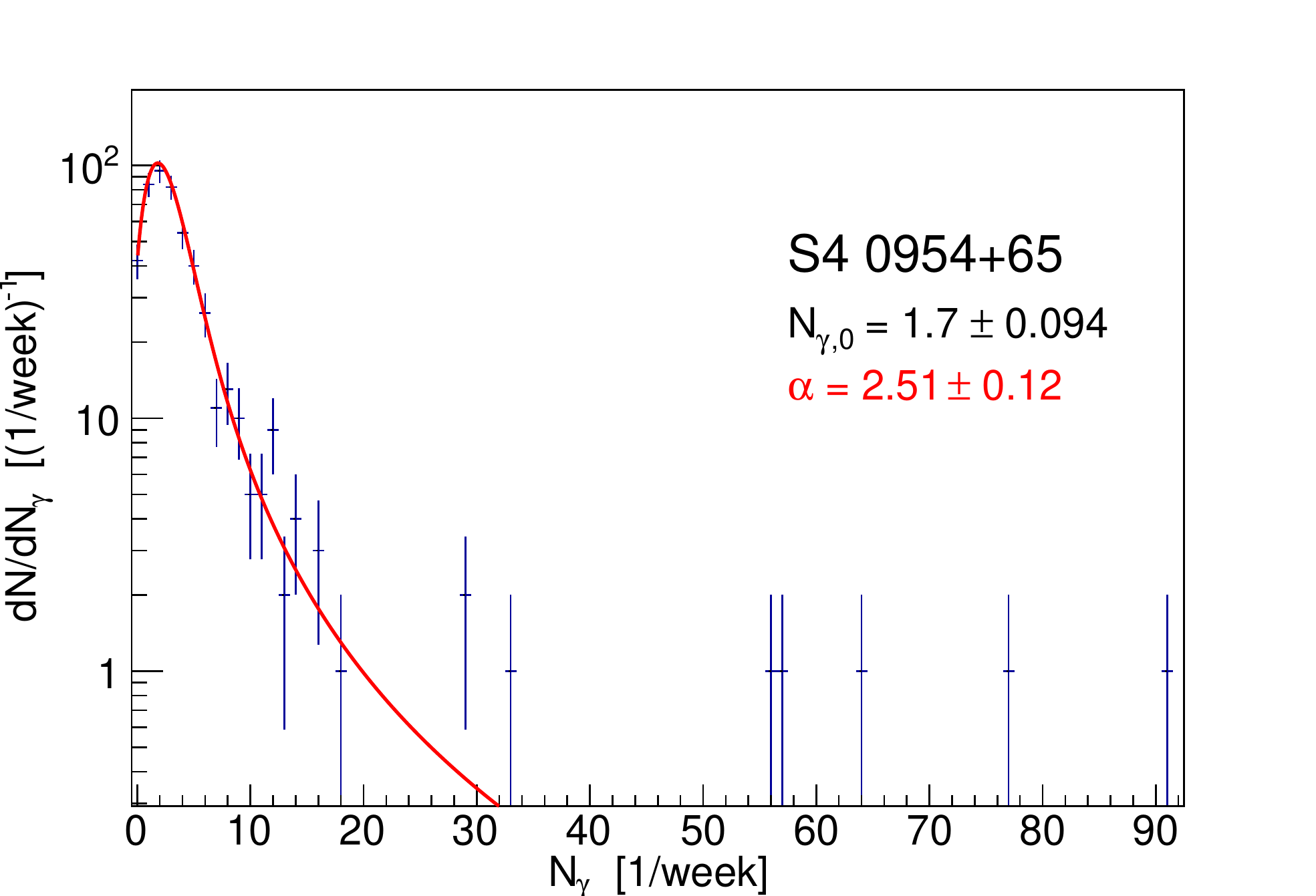}
\vspace{-0.4cm}
\caption{Histogram of the number of photons, $N_{\gamma}$, detected per week by the FAVA analysis in the direction of TXS 0506+056, OJ\,287, PKS\,0301-243, and S4\,0954+65 in the high-energy bin ($800$~MeV$-$300~GeV). The photon distribution is modelled as a power law with spectral index $\alpha$ convolved with a Poissonian distribution (solid red  line).}
\label{fig:fig1}      
\vspace{-0.4cm}
\end{figure*}
In addition, we quantify the fraction of energy emitted in the flaring state, $b_{\rm fl}$, defined as,
\begin{equation}
\label{eq:bfl}
b_{\rm fl}=\frac{1}{L^{\rm ave}N_{\rm tot}}\int_{L^{\rm th}}\mathrm{d} L\, L\frac{\mathrm{d}N}{\mathrm{d}L},
\end{equation}
where the average luminosity is given by $L^{\rm ave}=(1/N_{\rm tot})\int \mathrm{d}L\, L(\mathrm{d}N/\mathrm{d}L)$ and the average flaring luminosity is $L^{\rm fl}=(b_{\rm fl}/f_{\rm fl})L^{\rm ave}$. In the flare-dominated limit, where, $b_{\rm fl}\approx 1$, $L^{\rm ave}$ approaches $L^{\rm ave}\approx f_{\rm fl}L^{\rm fl}$.    
We determined $b_{\rm fl}$ and $f_{\rm fl}$ for a sample of FAVA sources by using the weekly-binned publicly available photon counts for each source. We thus obtain the photon-count distribution, ${\rm d}N/{\rm d} N_{\gamma}$, which is proportional to the \gammarayAdj luminosity distribution, ${\rm d}N/{\rm d} L_{\gamma}$, of the source in this time and energy bin, as long as the photon index doesn't change. For the rest of this discussion we assume ${\rm d}N/{\rm d} N_{\gamma} \sim {\rm d}N/{\rm d} L_{\gamma}$. 

Table \ref{tab:1} presents our results for a small sample of sources at redshifts similar 
to that of TXS\,0506+056. We find that the duty factor lies in the range of $0.3-10$\% for $\geq5\sigma$ flares 
(where for the significance quoted we use the FAVA definition), and obtain $f_{\rm fl}\approx0.02-0.1$ for TXS 0506+056. 
The corresponding fraction of emitted photons is $b_{\rm fl}\sim0.1\ll1$, 
implying that the bulk of the $\gamma$-ray emission of the studied sources comes from quiescent periods. 
\begin{table}[t]
\caption{Flare duty factor, $f_{\rm fl}$, and fraction of energy released during flares, $b_{\rm fl}$, for a sample of sources as derived from the FAVA analysis. We report values for the low-energy, {\it LE}, $100-800$~MeV, and high-energy, {\it HE}, $800$~MeV$-300$~GeV, bins. The duty factors quoted are for flux enhancement $\geq 5\sigma$ according to the FAVA definition. The $\gamma$-ray luminosity of the sources, $L_{\gamma}$ (in units of ${\rm erg}~{\rm s}^{-1}$), is derived from the 3FGL in the $1-100$~GeV energy range.
Rounded values of the power-law index, $\alpha$, are also shown. 
}
\label{tab:1}
\vspace{-0.4cm}
\begin{center}
\small
\begin{tabular}{|c||c|c|c|c|c|c|}
\hline 
Name & $L_{\gamma}$ & $f_{\rm fl}^{\rm LE}$ & $f_{\rm fl}^{\rm HE}$ & $b_{\rm fl}^{\rm LE}$ & $b_{\rm fl}^{\rm HE}$ & $\alpha$\\
\hline 
TXS 0506+056 & ${10}^{46.3}$ & 0.1 & 0.03 & 0.1 & 0.1 & $3.0$\\
\hline
OJ 287 & ${10}^{46.1}$ & 0.02 & 0.01 & 0.04 & 0.1 & $2.9$ \\
\hline
PKS 0426-380 & ${10}^{48}$ & $0.1$ & $0.1$ & 0.2 & 0.2 & $1.7$\\
\hline
PKS 0301-243 & ${10}^{46}$ & $0.04$ & $0.05$ & 0.1 & 0.3 & $2.5$\\
\hline
S5 0716+071 & ${10}^{46.7}$ & 0.07 & 0.08 & 0.1 & 0.2 & $1.7$\\
\hline
S4 0954+065 & $10^{45.5}$ & 0.04 & 0.03 & 0.07 & 0.3 & $2.5$\\
\hline
\end{tabular}
\vspace{-0.7cm}
\end{center}
\end{table}
We further investigated the frequency and duty cycle of flares by fitting the photon count distribution of each FAVA source with a power-law, $ {\rm d}N/{\rm d} N_{\gamma} \sim {\rm d}N/{\rm d} L_{\gamma} \propto  L_{\gamma}^{-\alpha}$. The number of detected photons per time interval is then given by a convolution of this power law with a Poissonian distribution. Example fits for TXS\,0506+056 and some of the other sources analysed are shown in Fig.~\ref{fig:fig1}. For TXS\,0506+056, $\alpha \simeq 3$. For the other sources analysed we found $\alpha \sim 2-4$. 
The above finding that the duty factor of flares and corresponding fraction of emission during flaring is subdominant might lead to the conclusion that the majority of neutrinos are produced during quiescent states in blazars. However, even though for the sources that we studied
highly-significant flaring states are achieved for a small fraction of time, during which a non-dominant part of the observed \gammarayAdj emission is released this may not be the case for the neutrino emission for at least two reasons, which we discuss below in turn. 

Firstly, neutrinos can be preferentially produced during $\gamma$-ray flares as typically, in models where the majority of $\gamma$-rays are leptonic in origin, $L_{\nu} \propto L_{\rm rad}^{\gamma}$, with $\gamma \sim 1.5 - 2.0$. This was demonstrated, for example, in \cite{Petropoulou:2016ujj, Murase:2016gly, Tavecchio:2014xha} and references therein. Therefore,  
\be L_{\nu}^2 \frac{{\rm d}N}{{\rm d}L_{\nu}} \propto L_{\nu}^{ 1 -
	    [(\alpha - 1)/\gamma]},
\ee	
which implies that neutrino production from flaring states can dominate the total neutrino emission of such sources even if flares don't dominate the radiative output. This is, for example, the case for sources with $\gamma \sim 2.0$ and $\alpha \leq 3$. As shown in fig.~\ref{fig:fig1} the FAVA data of TXS\,0506+056 and some of the other sources we analysed are consistent with these sources being in the latter category. 

Second, neutrino emission can be significantly enhanced during flaring, if the proton spectrum changes to a harder spectral index. In order to demonstrate this, we consider as an example a source where protons are produced with a power-law spectrum with index $s_{\rm l} = 2.3$ during steady state emission, and a proton spectrum which becomes harder during the flare, with index, $s_{\rm fl} = 1.8$. For such a scenario we find that the enhancement factor $c[\vareps
_{\nu}]$ is given by,  
\be 
c[\vareps
_{\nu}] =  \frac{\varepsilon_{\nu}L^{\rm fl}_{\varepsilon_{\nu}}}{\varepsilon_{\nu}L^{\rm l}_{\varepsilon_{l}}} \approx \frac{(2-s_{\rm fl})}{(2-s_{\rm l})}\frac{f_{p\gamma}^{\rm fl}}{f_{p\gamma}^{\rm l}} \left( \frac{20 \varepsilon_{\nu}}{\varepsilon_{p_{\rm max}}^{\rm fl}}{} \right)^{2-s_{\rm fl}}
\left( \frac{20 \varepsilon_{\nu}}{\varepsilon_{p_{\rm max}}^{\rm l}}{} \right)^{s_{\rm l}-2}.
\ee
\noindent If we further assume that the neutrino energy is $\vareps_{\nu} = 0.05 \vareps_{{\rm fl} p,{\rm max}} = 0.1$~PeV, and the minimum proton energy during steady state is $\vareps_l^{\rm fl} = 10$~GeV, we find $c[\vareps_{\nu}] \sim 30  f_{p\gamma}^{\rm fl}/f_{p\gamma}^{\rm l}$, which further demonstrates that it is possible that neutrino emission during electromagnetic flares dominates the total neutrino output of the source. 

We conclude that the contribution of flaring blazars like those observed from TXS 0506+056 to the diffuse background can be limited as,
\ba E_{\nu}^2 \Phi_{\rm \nu} & 
\lesssim 3.8 \times 10^{-10}~{\rm
 GeV}~{\rm cm}^{-2}~{\rm s}^{-1}~{\rm sr}^{-1} \nonumber \\
 & \left( 
 \frac{2\pi}{\Delta \Omega}\right) \left(\frac{\xi_z}{0.7} \right)
 q_L^{-1} \left( \frac{0.05}{f_{\rm fl}}
 \right)^{1/2} \nonumber \\ & \times \left( \frac{10^{46}~{\rm erg}~{\rm
 s}^{-1}}{\varepsilon_{\nu}L_{\varepsilon_{\nu}}} \right)^{1/2}
 F_{\rm lim,-9.5}^{3/2}.
 \ea
Note that the general contribution cannot exceed the limit of equation \ref{eq:upperLim}.  
\begin{figure}
\centering
\includegraphics[width=8cm,clip]{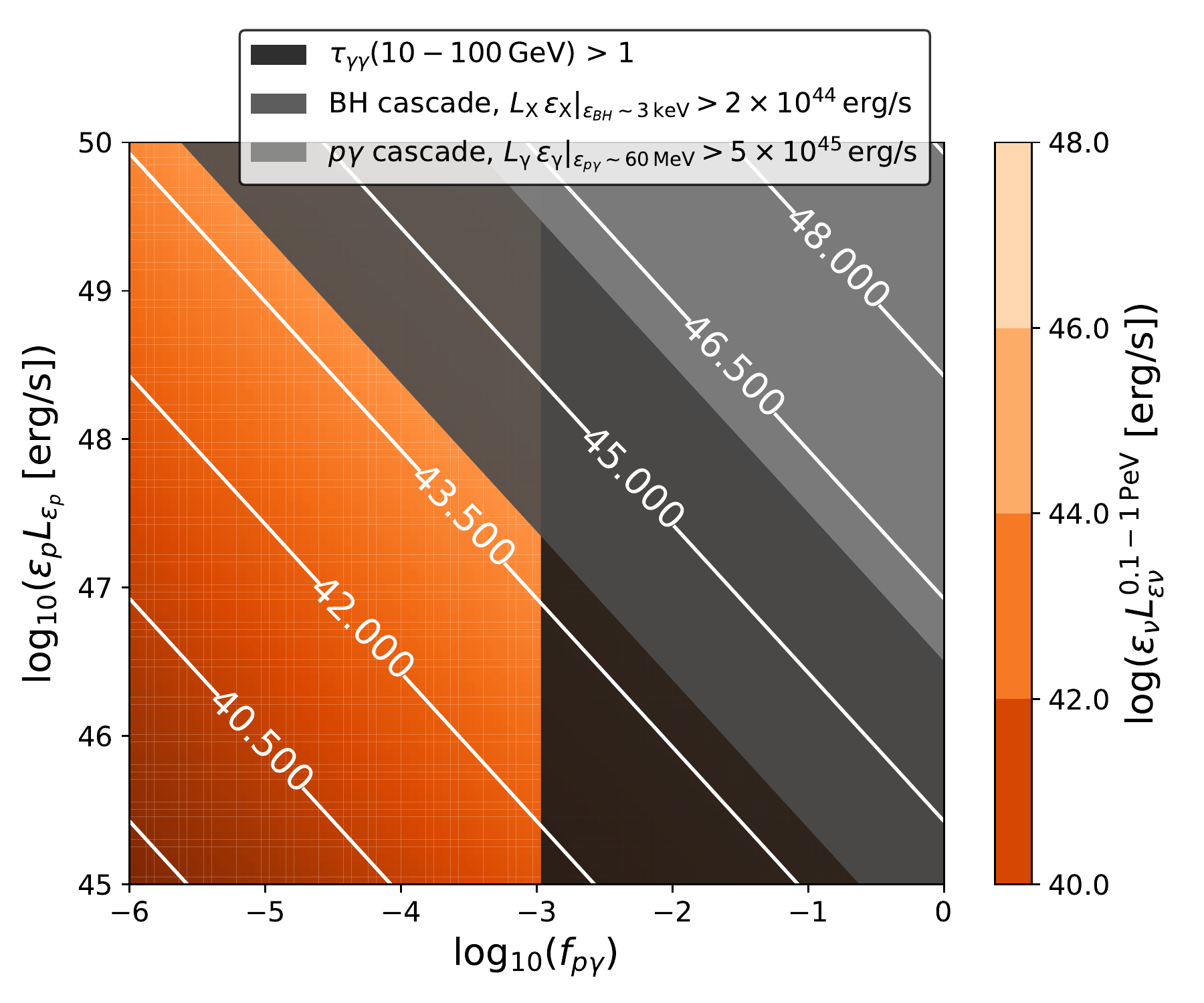}
\vspace{-0.4cm}
\caption{Maximum all-flavor neutrino luminosity, $\varepsilon_{\nu} L_{\varepsilon_{\nu}}^{\rm 0.1 - 1.0 \,PeV}$, (white contour lines), in the $\sim 0.1 - 1$~PeV energy range as a
  function of the injected proton luminosity, $\varepsilon_p L_{\varepsilon_{p}}$, and the neutrino
  production efficiency from \pgamma interactions, \fpg, in a
  single-zone model for the 2017 flare of \TXS, for protons with energy $\varepsilon_p \sim 20 \, \varepsilon_{\nu}$. The region where \fpg $>10^{-3}$ (black) is
  excluded by the observation of $> 100$~GeV photons from
  TXS\,0506+056 during the 2017 flare. Mid-dark grey shows the region 
  ruled out by the X-ray cascade constraint (eq. \ref{eq:BHconst}) and the lighter 
 grey by the \gammarayAdj cascade constraint (eq. \ref{eq:pgconst}).}
\label{fig:constraints}
\vspace{-0.4cm}
\end{figure}

\subsection{Constraints from X-ray and $\gamma$-ray observations}
\label{subsec:sec-2}
In single-zone blazar models neutrinos and $\gamma$-rays are coproduced 
inside the blazar blob through interactions with the photons in the blob. 
The optical depth for protons to \pgamma interactions, \fpg, is given by \cite{Murase:2015xka}, 
\be
f_{p \gamma} (\vareps_p) \approx \eta_{p\gamma}(\beta) \hat{\sigma}_{p\gamma} 
r'_b  n_{\varepsilon'_t} \varepsilon'_t|_{\vareps'_{t} = 0.5\bar{\varepsilon_\Delta}m_pc^2/ \vareps'_p}.
\ee
Here, $\eta_{p\gamma}(\beta)$ is an integration constant $\eta_{p\gamma}(\beta) \sim 2/(1+\beta)$, 
the effective \pgamma cross-section at threshold, $\hat{\sigma}_{p \gamma} \sim 0.7 \times 10^{-28}~{\rm cm}^2$,
$r'_b$ is the comoving radius of the emitting blob, $n_{\vareps'_t}$ is the density of target photons of energy 
$\vareps'_t$ and $\vareps'_p$ the comoving proton energy. 
Further, $m_p$ is the proton mass, and $m_{\pi}$ the pion mass. 
For \TXS it was shown in \cite{Keivani:2018rnh} that $\beta \sim 2.8$.  
\begin{figure}
\centering
\includegraphics[width=8cm,clip]{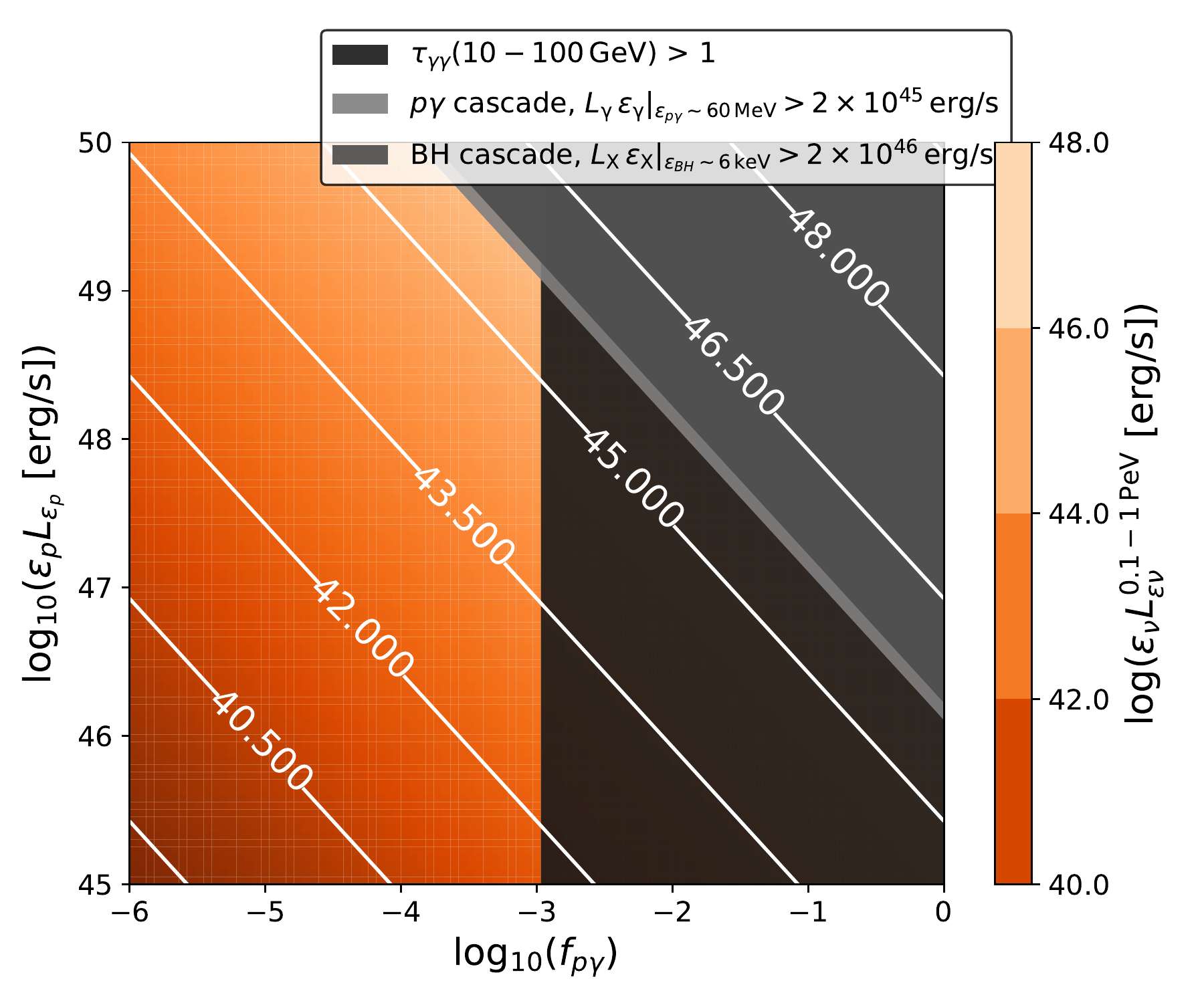}
\vspace{-0.4cm}
\caption{Same as figure \ref{fig:constraints} but for the 2014-15 IceCube archival flare.}
\label{fig:constraintsArch}
\vspace{-0.4cm}
\end{figure}
The same photons are the target photons for $\gamma \gamma$
interactions. The relevant optical depth is related to
the photo-meson production efficiency, $f_{ p\gamma}$, via,
\be
\tau_{\gamma \gamma} (\varepsilon_{\rm \gamma}) 
\approx \eta_{\gamma \gamma}(\beta) r'_{\rm b} \sigma_{\gamma \gamma} \vareps'_t n_{\vareps'_t} |_{\vareps'_t = m_e^2 c^4/\vareps'_{\gamma}}. 
\ee
\noindent with $\sigma_{\gamma \gamma} \sim 0.1\sigma_{\rm T}\sim 10^{-25}~{\rm cm}^2$ the Thompson cross-section  and $\eta_{\gamma \gamma}$ the integration constant of order unity. 
The ratio of optical depths from the two processes is then, 
\ba
f_{p\gamma} (\vareps_p) \approx \frac{\eta_{p \gamma}(\beta)}{\eta_{\gamma \gamma}(\beta)} \frac{\sigma_{\pgamma }}{\sigma_{\gamma \gamma}} \tau_{\gamma \gamma} (\vareps_{\gamma})
\approx 10^{-3} \tau_{\gamma \gamma}(\vareps_{\gamma}),
\ea
\noindent at energy, 
\be
\varepsilon_{\gamma} \sim {15~\rm GeV} ~ \left(\frac{\varepsilon_p}{6~{\rm PeV}}\right) 
\sim {15 ~\rm GeV} ~ \left(\frac{\varepsilon_{\nu}}{300~{\rm TeV}}\right).
\ee
\noindent Additionally, the same protons undergo Bethe-Heitler interactions \cite{Murase:2010va,petro_2015}, with effective optical depth, 
\ba
f_{BH} (\vareps_p) & \approx & \eta_{p\gamma}(\beta) \hat{\sigma}_{\rm BH} 
r'_b  n_{\varepsilon'_t} \varepsilon'_t |_{\vareps'_{t} = m_{p}c^2 \bar{\vareps}_{\rm BH} /2  \vareps'_p} \\
&=&g[\beta]f_{p\gamma}[\varepsilon_p],
\ea
\noindent where,  $\hat{\sigma}_{\rm BH}\sim0.8\times{10}^{-30}~{\rm cm}^2$, $g[\beta]\sim0.011{(30)}^{\beta-1}$, and $\bar{\vareps}_{\rm BH}  \sim 10(2m_e c^2) \sim 10 $ MeV.  
\noindent The observation of $>$10-100 GeV photons from \TXS during the
2017 flare, implies that the optical depth for photons to \gg
interactions on low-energy photons, $\tau_{\gg} (10-100~{\rm GeV}) <
1$. 

For proton luminosity, $\varepsilon_{p}L_{\varepsilon_p}$, the differential neutrino luminosity is then given by,
\begin{eqnarray}
\varepsilon_{\nu}L_{\varepsilon_\nu}&\approx&\frac{3}{8}f_{p \gamma}(\varepsilon_{p}L_{\varepsilon_p}) \,\,\,\,\,\,\,\,\,\,\,\,\,\,\,\,\,\,\,\,\,\,\,\,\,\,\,\,\,\,\,\,\,\,\,\,\,\,\,\,\,\,\,\,\,\,\,\,\,\,\,\,\,\,\,\,\,\,\,\,\,\,\,\,\,\,\,\,\,\,\,\,\,\,\,\,\,\,
\nonumber\\
&\simeq&1.2\times{10}^{45}~{\rm erg}~{\rm s}^{-1}~\frac{f_{p\gamma}}{10^{-4}}\left(\frac{\varepsilon_{p}L_{\varepsilon_p}}{{10}^{49.5}~{\rm erg}~{\rm s}^{-1}}\right). 
\label{eq:nuFlux}
\end{eqnarray}
\noindent We plot the relation given by eq. \ref{eq:nuFlux}, in fig.~\ref{fig:constraints}, 
where $\varepsilon_{\nu}L_{\varepsilon_\nu}$ is given by the orange color-map, and  white contour lines.
 
The constraint  $\tau_{\gg} (10-100~{\rm GeV}) <1$
imposes a limit to $f_{p \gamma}$, and thus to the maximum neutrino luminosity, $L_{\nu}$, in the one-zone
scenario. We therefore plot the region where \fpg $> 10^{-3}$ as excluded in fig.~\ref{fig:constraints}. 

A further limit is imposed on the maximum neutrino luminosity 
from the requirement that the synchrotron cascade flux produced by electron-positron pairs, injected by 
Bethe-Heitler interactions, should not exceed the observed X-ray flux. 
\ba
\varepsilon_{\gamma} L_{\varepsilon_{\gamma}}^{X}|_{\varepsilon_{\rm syn}^{\rm BH}} 
&\approx& \frac{1}{2(1+Y_{\rm IC})}g[\beta] f_{p\gamma}\,\varepsilon_{p} L_{p}\nonumber \\
&\approx& \frac{4}{3(1+Y_{\rm IC})}g[\beta] \,\varepsilon_{\nu} L_{\varepsilon_{\nu}},
\label{eq:BHconst}
\ea
\noindent where, $\varepsilon_{\rm syn}^{\rm BH} \approx 6~{\rm keV} B^\prime_{0.5~{\rm G}}(\varepsilon_{p}/6~{\rm PeV})^2(20/\delta)$, with $\delta$ the Doppler factor of the relativistic motion of the emitting region, $B'$ the magnetic field strength as measured by a comoving observer, and $Y_{\rm IC}$ the Compton dominance parameter which is at most 1 for \TXS \cite{Keivani:2018rnh}. For the flaring spectrum of \TXS from the analysis of \cite{Keivani:2018rnh} this corresponds to $ \varepsilon_\gamma L_{\varepsilon_\gamma}^X \leq 3 \times 10^{44}~{\rm erg}/{\rm s}$.

This imposes a constraint on the maximum muon neutrino luminosity at the level of, 
$\varepsilon_\nu L_{\varepsilon_{\nu_\mu}}^{0.1-1~{\rm PeV}}\lesssim \varepsilon_\gamma L_{\varepsilon_\gamma}^X/3 \sim{10}^{44}~{\rm erg}~{\rm s}^{-1}$, where the factor of 3 accounts for going from all-flavour to single-flavour neutrino luminosity. Eq. \ref{eq:BHconst} is plotted as the mid-grey exclusion region in fig.~\ref{fig:constraints} assuming $Y_{\rm IC} = 1$ which gives the most optimistic estimate for the maximum neutrino luminosity of \TXS.

In addition, a synchrotron cascade flux component from electron-positron pairs injected from photo-meson production and electron-positron pair production from hadronic $\gamma$ rays is unavoidable in the single-zone model. The minimum cascade flux from these processes is,
\be
\varepsilon_{\gamma}L_{\varepsilon_\gamma}|_{\varepsilon_{\rm syn }^{p\gamma}} \approx \frac{5}{8} \frac{1}{2( 1+Y_{\rm IC})}f_{p\gamma}(\varepsilon_{p}L_{\varepsilon_p}) \approx \frac{5}{6(1+Y_{\rm IC})}\varepsilon_{\nu}L_{\varepsilon_\nu},
\label{eq:pgconst}
\ee
where $\varepsilon_{\rm syn}^{p\gamma}\simeq60~{\rm MeV}~(B'/0.3~{\rm G)}){(\varepsilon_p/6~{\rm PeV})}^2(20/\delta)$.
The constraint of eq.~\ref{eq:pgconst} is shown as a further exclusion region in fig.~\ref{fig:constraints}, where we conservatively assume that $\varepsilon_{\gamma}L_{\varepsilon_\gamma}|_{\varepsilon_{\rm syn }^{p\gamma} \sim 60~\rm MeV} \leq 5 \times 10^{45}\rm erg s^{-1}$. As evident, for the 2017 flare of \TXS, the X-ray cascade constraint of eq.~\ref{eq:BHconst} is stronger.

Fig.~\ref{fig:constraintsArch} shows the cascade constraints for the archival, 2014-15-neutrino flare. Here, we plot exclusion regions for 
$\varepsilon_{\gamma}L_{\varepsilon_\gamma}|_{\varepsilon_{\rm syn }^{p\gamma} \sim 60~\rm MeV} \leq 2 \times 10^{45}\rm erg s^{-1}$, and 
$\varepsilon_{\gamma}L_{\varepsilon_\gamma}|_{\varepsilon_{\rm syn }^{p\gamma} \sim 10~\rm keV} \leq 2 \times 10^{46}\rm erg s^{-1}$, based on the analyses of \cite{Rodrigues:2018tku,Reimer:2018vvw}. In this case the~MeV data provide a stronger constraint. In deriving constraints in the MeV region for fig. \ref{fig:constraintsArch} and \ref{fig:constraintsArch} we have assumed that there is no additional, Bethe-Heitler bump in the SED (see e.g fig. 1 of \cite{Rodrigues:2018tku}). If such a component did exist, our upper limit for the 2014-15 flare would be too conservative.

Equations \ref{eq:BHconst} and \ref{eq:pgconst} show that the luminosity of the synchrotron cascade is comparable to the neutrino luminosity in single-zone models. Thus, the 2017 and 2014-15 flares reported by the IceCube Collaboration should be accompanied by X-ray emission with 
\be
E_{\gamma} F_{E_{\gamma}}^{X} \sim E_{\nu} F_{E_{\nu}}^{\rm 0.1-1 PeV} \sim (3-30) \times 10^{-11}~{\rm erg}~{\rm cm^{-2}}~{\rm s^{-1}},
\ee
\noindent which should be detectable by X-ray monitoring instruments such as {\it Swift} and {\it MAXI}. 

\section{Multi-zone model and cosmic-ray induced neutral beams}
\label{sec:sec-3}
The electromagnetic cascade emission discussed in the previous section
is a consequence of energy conservation, and hence expected not only in 
single-zone models, but also in multi-zone models. It is, therefore, crucial to take into account cascades inside the source. 

In this section we discuss the cosmic-ray induced neutral beam model, which was previously studied in \cite{Murase:2011yw} and \cite{Dermer:2012rg} as a possible way to explain the 2017 and 2014-15 flare of \TXS \, in a ``common'' framework. 
It has the appealing feature that it can avoid the cascade constraints, as isotropisation of high energy electrons and positrons is expected to take place if such conditions are found in the source environment. 
\begin{figure}
\centering
\sidecaption
\includegraphics[width=6cm,clip]{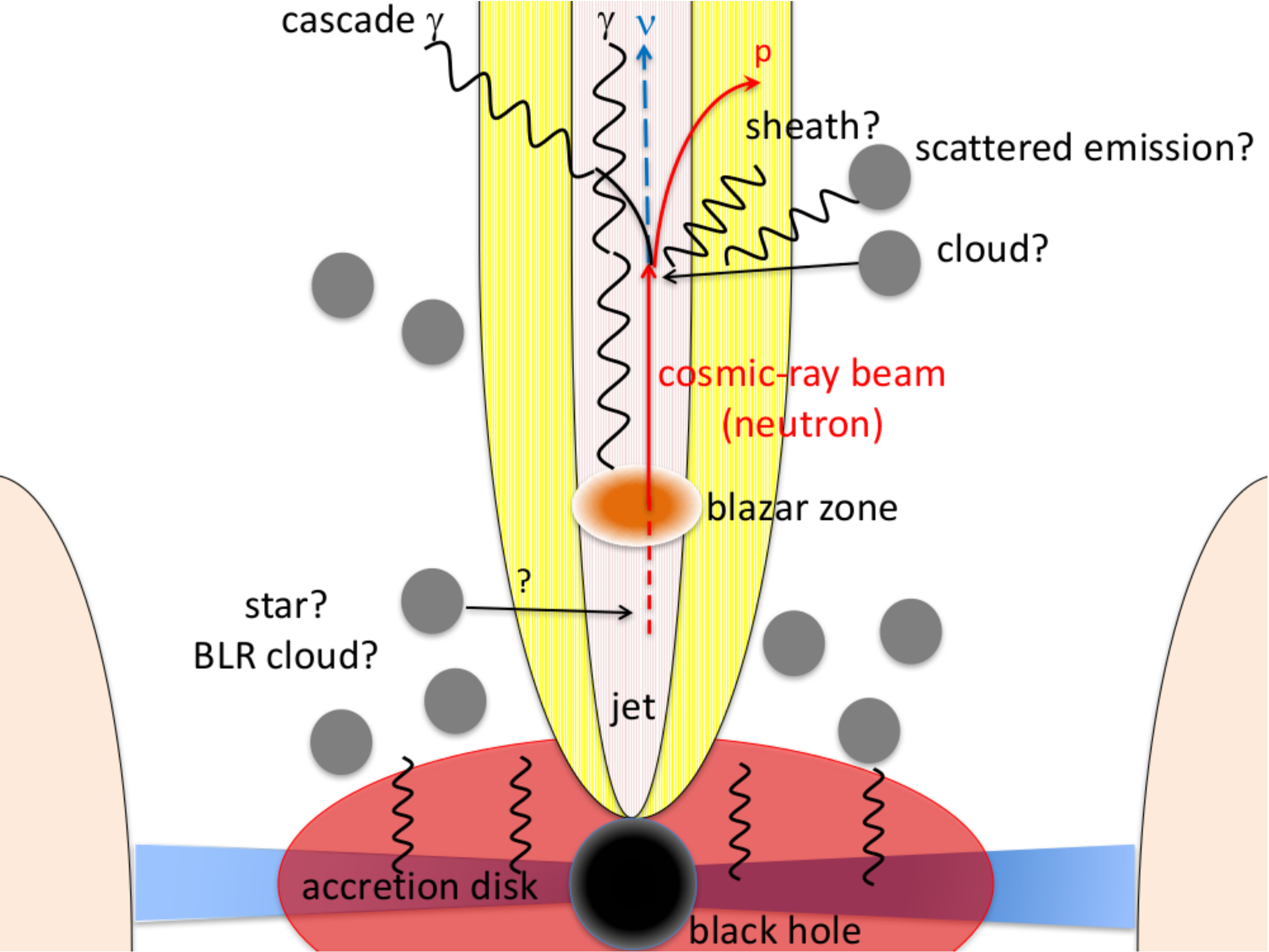}
\caption{Schematic representation of the cosmic-ray beam model for
  high-energy neutrino production. While the neutrino emission is
  highly beamed the associated cascade emission in the X-ray range is
  isotropised.}
\label{fig:sketch} 
\vspace{-0.4cm}
\end{figure}

For the cosmic-ray induced neutral beam model to be in operation, 
nuclei must be accelerated in the emitting region of the blazar. 
Subsequently, the following steps take place:
\begin{enumerate}
\item Neutrons are produced via the photo-disintegration of
nuclei in the cosmic-ray acceleration region. This region can be the vicinity of the central black hole. 
\item Nuclei and protons remain confined and eventually cool via adiabatic losses while
neutrons escape the cosmic-ray acceleration zone.
\item The neutrons keep to interact with external radiation fields (e.g., scattered accretion disk emission and non-thermal emission from the sheath) or dense clouds that
could exist at larger radii producing neutrinos.
\item The relativistic pairs
produced in $n\gamma$ or $np$ interactions get isotropised by the
magnetic 
fields in the jet or the surrounding medium. The emitted synchrotron radiation is not boosted, thus suppressing the
electromagnetic cascade that is otherwise expected. 
\end{enumerate}

\noindent A schematic representation of the model is shown in fig.~\ref{fig:sketch}.

The electromagnetic cascade is doubly suppressed since the electron-positron pairs can be isotropised in the larger scale jet, which also causes a spread in time, and in addition, neutrons don't undergo Bethe-Heitler interactions. The expected neutrino flux in this model is,
\be
\varepsilon_{\nu} L_{\varepsilon_{\nu}} \approx 3.8 \times 10^{46}~{\rm erg/s}
 \left(\frac{2K}{1+K} \right) \left(\frac{f_{n\gamma/np}\varepsilon_{n} 
 L_{\varepsilon_{n}}}{10^{47}~{\rm erg/s}}\right),
\label{eq:multizonenuL}
\ee
where, $f_{n\gamma/np}$ is the neutrino production efficiency in the interactions of neutrons with photons/hadrons respectively, 
$K = 1$ for $np$ and $K = 2$ for $n\gamma$ interactions. 
This model can therefore explain the energetics of both the 2017 and 2014-15 neutrino flares of \TXS as long as the neutrino production efficiency is $> 0.1$ and $\sim 1$, respectively. 
The neutron luminosity quoted in eq.~\ref{eq:multizonenuL} can be produced via photo-nuclear interactions in the blazar zone as detailed in~\cite{Murase:2018iyl}. 
\section{Conclusions}
\label{sec:sec-4}
In light of the recent finding of a high-energy neutrino in the direction of the flaring \TXS, we presented the summary of constraints on blazars as sources of the diffuse neutrino flux seen by IceCube, and specifically the constraints on \TXS, and by extension, single powerful blazars as neutrino sources. Our results reiterate that it is difficult to reconcile blazars with all existing neutrino observations as the dominant source population. We quantified the maximum neutrino flux that can be produced by individual blazars, and showed that this is limited by the requirement that the cascade flux produced by photons co-produced in interactions of hadrons inside the source, should not exceed multi-wavelength contemporaneous observations. We presented a multi-zone model based on the concept of cosmic-ray induced neutral beams, which can escape the X-ray and \gammarayAdj constraints that otherwise limit the maximum attainable neutrino luminosity from \TXS. 

\bibliography{biblio}
\end{document}